\documentclass[aps,reprint,twocolumn,floatfix,showpacs,pra,,superscriptaddress]{revtex4-1}
\usepackage[colorlinks,linkcolor=blue,anchorcolor=blue,citecolor=blue,filecolor=blue,menucolor=blue,runcolor=blue,urlcolor=blue,frenchlinks=blue]{hyperref}
\usepackage{graphicx}
\usepackage{subfigure}
\usepackage{amsmath}
\usepackage{amssymb}
\usepackage{dcolumn}
\usepackage{mathrsfs}
\usepackage{bm,physics}
\usepackage{color,amsfonts,amsthm,mathtools}
\usepackage{subfigure}
\usepackage{setspace}
\usepackage{bm}

\makeatletter

\newcommand{\Rmnum}[1]{\expandafter\@slowromancap\romannumeral #1@}
\makeatother

\begin{document}
\title{Synthetic non-Abelian topological charges in ultracold atomic gases}

\author{Qi-Dong Wang}
\affiliation{National Laboratory of Solid State Microstructures and School of Physics, Nanjing University, Nanjing 210093, China}
\affiliation{Collaborative Innovation Center of Advanced Microstructures, Nanjing 210093, China}
\affiliation{Key Laboratory of Atomic and Subatomic Structure and Quantum Control (Ministry of Education), Guangdong Basic Research Center of Excellence for Structure and Fundamental Interactions of Matter, School of Physics, South China Normal University, Guangzhou 510006, China}
\affiliation{Guangdong Provincial Key Laboratory of Quantum Engineering and Quantum Materials, Guangdong-Hong Kong Joint Laboratory of Quantum Matter, Frontier Research Institute for Physics, South China Normal University, Guangzhou 510006, China}

\author{Yan-Qing Zhu}
\affiliation{Guangdong-Hong Kong Joint Laboratory of Quantum Matter, Department of Physics, and HK Institute of Quantum Science \& Technology, The University of Hong Kong, Pokfulam Road, Hong Kong, China}
\affiliation {Quantum Science Center of Guangdong-Hong Kong-Macao Greater Bay Area, Shenzhen, China}

\author{Shi-Liang Zhu}\email{slzhu@scnu.edu.cn}
\affiliation{Key Laboratory of Atomic and Subatomic Structure and Quantum Control (Ministry of Education), Guangdong Basic Research Center of Excellence for Structure and Fundamental Interactions of Matter, School of Physics, South China Normal University, Guangzhou 510006, China}
\affiliation{Guangdong Provincial Key Laboratory of Quantum Engineering and Quantum Materials, Guangdong-Hong Kong Joint Laboratory of Quantum Matter, Frontier Research Institute for Physics, South China Normal University, Guangzhou 510006, China}
\affiliation {Quantum Science Center of Guangdong-Hong Kong-Macao Greater Bay Area, Shenzhen, China}

\author{Zhen Zheng}\email{zhenzhen@m.scnu.edu.cn}
\affiliation{Key Laboratory of Atomic and Subatomic Structure and Quantum Control (Ministry of Education), Guangdong Basic Research Center of Excellence for Structure and Fundamental Interactions of Matter, School of Physics, South China Normal University, Guangzhou 510006, China}
\affiliation{Guangdong Provincial Key Laboratory of Quantum Engineering and Quantum Materials, Guangdong-Hong Kong Joint Laboratory of Quantum Matter, Frontier Research Institute for Physics, South China Normal University, Guangzhou 510006, China}

% Nanjing address:
%\affiliation{National Laboratory of Solid State Microstructures and School of Physics, Nanjing University, Nanjing 210093, China}
%\affiliation{Collaborative Innovation Center of Advanced Microstructures, Nanjing 210093, China}

% Hong Kong address:
%\affiliation{Guangdong-Hong Kong Joint Laboratory of Quantum Matter, Department of Physics, and HK Institute of Quantum Science \& Technology, The University of Hong Kong, Pokfulam Road, Hong Kong, China}

% Guangdong address:
%\affiliation{Key Laboratory of Atomic and Subatomic Structure and Quantum Control (Ministry of Education), Guangdong Basic Research Center of Excellence for Structure and Fundamental Interactions of Matter, School of Physics, South China Normal University, Guangzhou 510006, China}
%\affiliation{Guangdong Provincial Key Laboratory of Quantum Engineering and Quantum Materials, Guangdong-Hong Kong Joint Laboratory of Quantum Matter, Frontier Research Institute for Physics, South China Normal University, Guangzhou 510006, China}
%\affiliation {Quantum Science Center of Guangdong-Hong Kong-Macao Greater Bay Area, Shenzhen, China}

%----------------------------------------------------------------------------------------
\begin{abstract}
	Topological phases associated with non-Abelian charges can exhibit a distinguished bulk-edge correspondence compared with Abelian phases, although elucidating this relationship remains challenging in traditional solid-state systems. In this paper, we propose a theoretical framework for synthesizing non-Abelian quaternion charges in ultracold atomic gases.
	By designing artificial spin-orbit coupling patterns, the topological edge modes demonstrate a clear correspondence with the band topology determined by various quaternion charges. This paves the way for observing the interface modes whose existence is attributed to the nonconservation multiplication relation, which is fundamental to non-Abelian charges. This scheme can be readily implemented using current ultracold atom techniques, offering a promising approach to explore the intriguing non-Abelian characteristics of the system.
\end{abstract}
\maketitle
%----------------------------------------------------------------------------------------
\section{Introduction}

Topological band theory has extracted the nature of nontrivial phases by the topological charges,
which have attracted tremendous interests of studies.
In most previous works, the topological charges that specify the phases belongs to the $Z_2$ or $Z$ classes \cite{Hasan2010Nov-topo-review,Qi2011Oct-topo-review}, both represented by the Abelian groups.
As the results, the emergence and degeneracy of the topological edge modes depend on the bulk topological charges, known as the bulk-edge correspondence.
On the other hand, recent works report the findings of nontrivial phases described by a totally different classes such as non-Abelian quaternion group \cite{Wu2019Sep-science-non-abelian-start,Bouhon2020Nov-non-abelian-charges-theo}.
It leads to a distinguished bulk-edge correspondence associated with the non-Abelian features,
bringing in rich behaviors of emergent edge modes.

In the previous works, engineering topological phases of non-Abelian charges requires the deliberate prepare of the nontrivial intrinsic fields.
Although the realization is still frustrated in conventional solid-state systems, the emulation using artificial quantum systems shows an alternative routine for investigating the non-Abelian topological charges \cite{Guo2021Jun-non-abelian-charges-in-photonics,Jiang2021Nov-charges-in-photonics,Li2023Oct-non-abelian-charges-in-cold-atoms,Jiang2021Nov-q16-charges,Slager2024Feb-non-abelian-charges-in-kagome-model}.
Notably in recent decades, ultracold atomic gases have been widely applied in studies of quantum simulations \cite{Bloch2012Apr-simulations-in-atoms,Gross2017Sep-simulations-in-atoms}.
This is because it can provide a reliable and clean platform for investigating a broad range of topological phases \cite{Cooper2019Mar-topo-review-in-atoms,Zhang2018Oct-topo-review-in-atoms},
by taking advantages of its highly artificial controllability and manipulations.
Successful achievements have been obtained in engineering a variety of artificial fields~\cite{Ruseckas2005,Lin2011Mar-soc-exp,Lin2009,Jaksch2003,Hamner2014Jun-soc-exp,FMei2012,LZTang2021,Dalibard2011Nov-optic-method-review,Goldman2014Nov-optic-method-review,DWZhang2016},
such as atomic spin Hall effect \cite{Beeler2013,SLZhu2006,XJLiu2007} and quantum anomalous Hall effect~\cite{Jotzu2014,LBShao2008,CJWu2008},  spin-orbit coupling (SOC) \cite{Galitski2013,Huang2016Jun-nat-phys-soc-exp,Wu2016Oct-soc-exp,Livi2016Nov-soc-exp,An2017Apr-soc-exp,Kolkowitz2017Feb-soc-exp,Sun2018Oct-soc-exp,Lv2021Sep-soc-exp,Aeppli2022Oct-soc-exp,DWZhang2012,Xiong2023Dec-soc-appli},
artificial magnetic fields \cite{Lin2009Dec-nature-magnetic-field,Aidelsburger2011Dec-magnetic-field,Jimenez-Garcia2012May-prl-magnetic-field,Struck2012May-prl-magnetic-field,Aidelsburger2013Oct-magnetic-field,Miyake2013Oct-magnetic-field,Celi2014Jan-prl-magnetic-field,YPWu2022,XShen2022,Mandal2016Aug-su3-gauge-field},
artificial non-Abelian gauge fields \cite{Sugawa2021Sep-non-abelian-field-exp,Li2022Nov-pra-non-abelian-field-exp,Lv2023Aug-pra-non-abelian-field-exp},
and nontrivial many-body interactions \cite{Venu2023Jan-p-wave-interaction,SLZhu2013,Wu2020Jan-non-trivial-pairing,Wu2023Apr-non-trivial-pairing}.
Therefore, this motivates us to search a potential scheme for realizing phases characterized by non-Abelian charges using ultracold atoms,
which can also offer a valid tool for exploring and studying the relative non-Abelian physics \cite{Qi2009May-non-abelian-appli,Ghosh2010Nov-non-abelian-appli,Ding2022Jan-pra-non-abelian-appli,You2022Jun-non-abelian-appli,Sun2022Sep-non-abelian-appli,Ding2024Apr-pra-non-abelian-appli,Huang2024May-pra-non-abelian-appli}.

An earlier work reports a scheme for engineering the quaternion $Q_8$ charges in a Floquet system \cite{Li2023Oct-non-abelian-charges-in-cold-atoms}.
By manipulating the temporal sequence of Hamiltonians, the designed Floquet topological insulator phase demonstrates the presence of non-Abelian charges, focusing just on the interface between two quaternion charges within the same conjugate class. In contrast with Floquet-based approaches \cite{Eckardt2017Mar-floquet-review}, in this paper, we present a proposal based on a stable system of ultracold atoms. Instead of employing Floquet engineering, we synthesize  quaternion charges by preparing  various  patterns of SOC, enabling the observation of interface modes arising from domain walls between quaternion charges, even those belonging to distinct conjugate classes. This exploration provides clear evidence of the nonconservation multiplication relation, offering a promising avenue for studying and characterizing non-Abelian physics.

This paper is organized as follows:
In Sec. \ref{sec:model}, we start with the model Hamiltonian and the scheme for engineering the quaternion charges.
Based on the scheme, in Sec. \ref{sec:interface-modes}, we show that the junction structure that connects the two quaternion charges can be proposed by introducing external fields with a spatial offset.
Such a setup supports the interface-mode whose behaviors are determined by the noncommunicative multiplication relation between different quaternion charges,
which is the core of the non-Abelian charges.
In Sec. \ref{sec:experiment} we discuss the details for the experimental realization in ultracold atoms.
Finally we conclude this paper in Sec. \ref{sec:conclusion}.

\section{Model Hamiltonian}\label{sec:model}

We consider the quantum gases confined in a one-dimensional (1D) optical lattice.
We choose three internal states of the atoms as the pseudospins $\sigma=A,B,C$.
The model Hamiltonian is composed of two parts
\begin{equation}
	\hat{H} = \hat{H}_0 + \hat{H}_{\rm SC} \,. \label{eq:h-start}
\end{equation}
The first part describes the kinetic term in accompany of the optical lattice,
\begin{equation}
	\hat{H}_0 = \int dx\sum_{\sigma} \psi_{\sigma}^\dag (x) \big[ -\frac{\nabla_x^2}{2m} +\Gamma_{\sigma} + V_{\rm OL}(x) \big] \psi_{\sigma}(x) \,. \label{eq:h-0-start}
\end{equation}
Here $\psi_\sigma$ is the annihilation operator of atoms with spin-$\sigma=A,B,C$.
$\Gamma_{\sigma}$ is the on-site potential.
$V_{\rm OL}(x) = V_{L}\sin^2(k_Lx)$ is the lattice potential, where the $V_L$ characterizes the trap depth, and $k_{L}=\pi/\lambda_L$ with $\lambda_L$ denoting the lattice constant.
We have set $\hbar=1$ for simplicity of notation.
The second part of Hamiltonian (\ref{eq:h-start}) describes the coupling between different spins,
\begin{equation}
	\hat{H}_{\rm SC} = \int dx\sum_{\sigma\neq\sigma'}
	\hat{\Omega}_{\sigma\sigma'}(x) \psi_{\sigma}^\dag(x) \psi_{\sigma'}(x) \,. \label{eq:h-c-start}
\end{equation}
Here $\hat{\Omega}_{\sigma\sigma'}(x)$ denotes coupling field associated with the following standing-wave mode,
\begin{equation}
	\hat{\Omega}_{\sigma\sigma'}(x) = i\Omega_{\sigma\sigma'}\sin(k_Lx) \,, \label{eq:coupling-pattern}
\end{equation}
where $\Omega_{\sigma\sigma'}$ characterizes its amplitude.

\begin{figure}[t]
	\centering
	\includegraphics[width=0.48\textwidth]{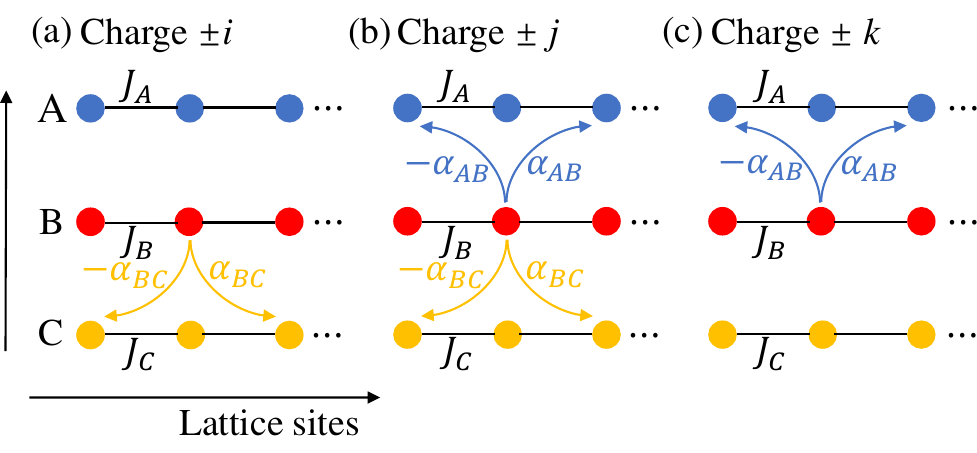}
	\caption{Setups of the 1D lattice models for engineering the quaternion charges (a) $\pm i$, (b) $\pm j$, and (c) $\pm k$.}
	\label{fig:single}
\end{figure}

We employ the tight-binding approximation to investigate such a lattice model.
Since atoms of all spins are loaded in the same lattice potential, we expand the atomic operator $\psi_{\sigma}(x)$ of different spins in terms of the same Wannier wave functions $W(x)$,
\begin{equation}
	\psi_{\sigma} = \sum_{j,\sigma} W(x-x_j) c_{j\sigma} \,.
\end{equation}
Here $c_{j\sigma}$ denotes the annihilation operator on the $j$-th site.
The profile of $W(x-x_j)$ is localized and symmetric with respect to each site center $x_j=j\lambda_L$.
Hamiltonians (\ref{eq:h-0-start}) and (\ref{eq:h-c-start}) are then recast as
\begin{align}
	H_0 &=\sum_{j,\sigma} \big[\Gamma_{\sigma} c_{j\sigma}^\dag c_{j\sigma} - \hat{J}_{\sigma} (c_{j+1,\sigma}^\dag c_{j\sigma}+\rm H.c.) \big] \,,\label{eq:h-0-site} \\
	H_{\rm SC} &= \sum_{j,j'}\sum_{\sigma\neq\sigma'} i\hat{A}_{jj'}^{\sigma\sigma'} c_{j\sigma}^\dag c_{j'\sigma'}+\rm H.c. \label{eq:h-c-site-0}
\end{align}
where $\hat{J}_{\sigma}$ denotes the magnitude of the nearest-neighbor (NN) tunneling,
\begin{equation}
	\hat{A}_{jj'}^{\sigma\sigma'} = \Omega_{\sigma\sigma'} \int \sin(k_Lx) W^*(x-x_j)W(x-x_{j'})dx \,,
\end{equation}
and $\rm H.c.$ stands for the Hermitian conjugate.
Due to the odd parity of the coupling pattern (\ref{eq:coupling-pattern}),
we can obtain $\hat{A}_{jj}^{\sigma\sigma'}=0$, i.e., the on-site coupling vanishes.
As the consequence, the NN term $\hat{A}_{j\pm1,j}^{\sigma\sigma'}$ is dominant.
Therefore, Eq. (\ref{eq:h-c-site-0}) is simplified as
\begin{equation}
	H_{\rm SC}= \sum_{j,\sigma\neq\sigma'} (-1)^ji\alpha_{\sigma\sigma'} (c_{j+1\sigma}^\dag c_{j\sigma'} - c_{j-1\sigma}^\dag c_{j\sigma'} )+\rm H.c. \label{eq:h-c-site}
\end{equation}
where we have represented $\hat{A}_{j+1,j}^{\sigma\sigma'}$ by $\hat{A}_{j+1,j}^{\sigma\sigma'} = (-1)^j \alpha_{\sigma\sigma'}$ with $\alpha_{\sigma\sigma'} = \Omega_{\sigma\sigma'} \int \sin(k_Lx) W^*(x-\lambda_L)W(x)dx$.
At this stage, one can find Eq. (\ref{eq:h-c-site}) reduces to a field that describes SOC \cite{Liu2014Feb-soc-theo}.

\begin{figure*}[t]
	\centering
	\includegraphics[width=0.9\textwidth]{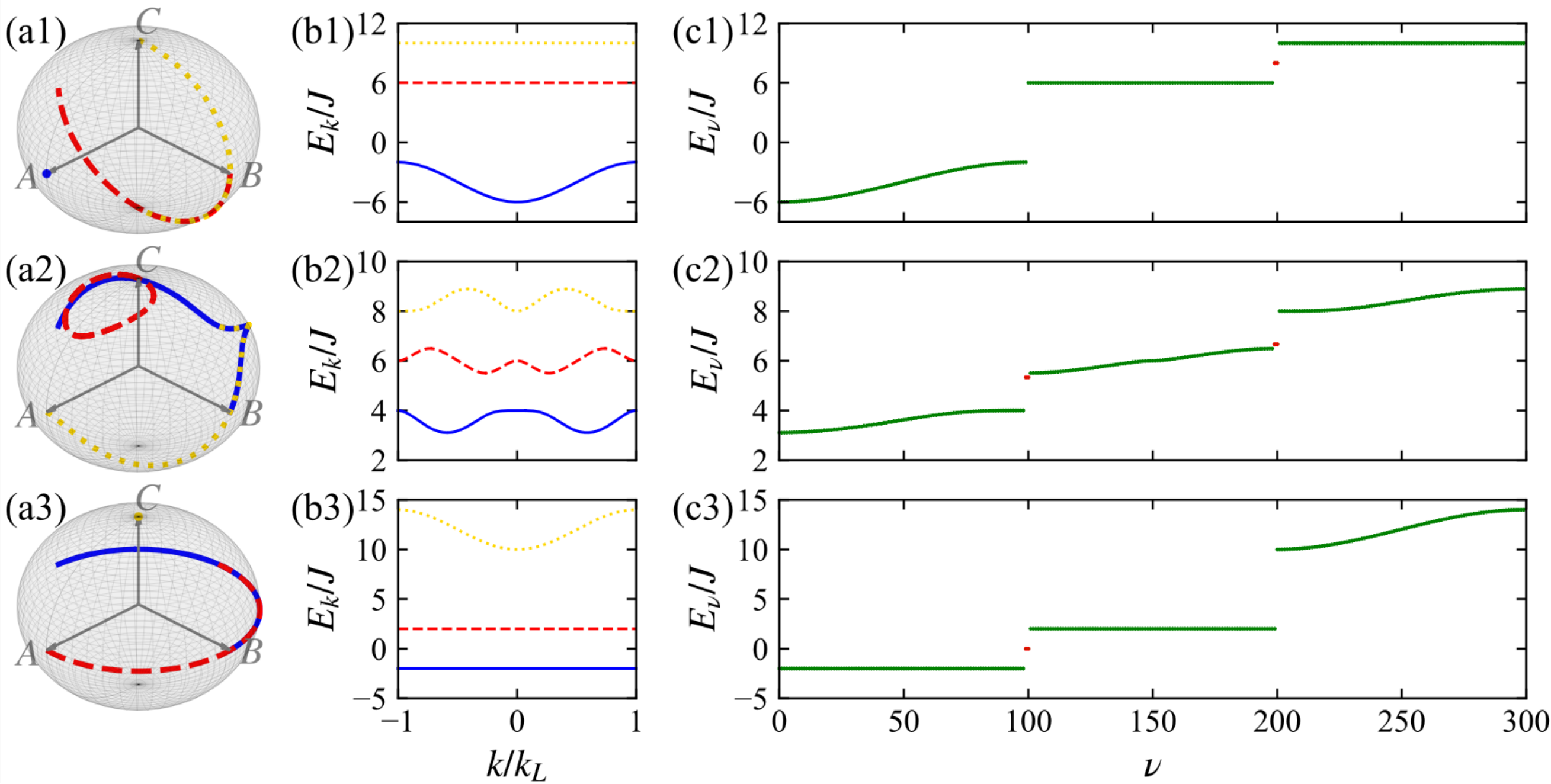}
	\caption{
		Band physics of the quaternion charges $i$ (top row), $j$ (center row), and $k$ (bottom row).
		(a1)-(a3) Evolution of the eigenstates for Hamiltonian (\ref{eq:h-k-space}) when $k$ swaps over BZ.
		The coordinates on the sphere stand for the components of spins $A$, $B$, and $C$.
		(b1)-(b3) Spectrum of Hamiltonian (\ref{eq:h-k-space}) in the momentum-$k$ space.
		The lowest, intermediate, and highest bands are respectively marked by the blue-solid, red-dashed, and yellow-dotted lines.
		(c1)-(c3) Spectrum of Hamiltonian (\ref{eq:h-single-charge}) in the real space.
		$\nu$ stands for the quasiparticle index.
		Topological edge modes are marked by red dots.
		We set $J_B=-J$ and use $J$ as the energy unit in the whole work.
		Other parameters are as follows.
		In the top row, we set $J_A=J_C=J$, $(\Gamma_A,\Gamma_B,\Gamma_C)=(-4J, 8J, 8J)$, and $(\alpha_{AB},\alpha_{CB})=(0,-J)$.
		In the center row, we set $(J_A,J_C)=(J,0)$, $(\Gamma_A,\Gamma_B,\Gamma_C)=(6J, 6J, 6J)$, and $(\alpha_{AB},\alpha_{CB})=(-J,J)$.
		In the bottom row, we set $J_A=J_C=J$, $(\Gamma_A,\Gamma_B,\Gamma_C)=(0, 0, 12J)$, and $(\alpha_{AB},\alpha_{CB})=(-J,0)$.
	}
	\label{fig:spectra of single charge}
\end{figure*}

For simplicity without loss of generality, we consider the couplings in Eq. (\ref{eq:h-c-site}) are only processed
from spin $B$ to spin $A$ or $C$.
Since the magnitude of SOC exhibits a staggered pattern in Eq. (\ref{eq:h-c-site}),
such a spatial modulation can be eliminated if one invokes the following operator representation
\begin{equation}
	c_{j,B} \rightarrow (-1)^{j} c_{j,B} \label{eq:operator-transformation}
\end{equation}
solely for spin $B$.
Under the transformation (\ref{eq:operator-transformation}),
the onsite term in Eq. (\ref{eq:h-0-site}) remains unchanged,
while the hopping term $\hat{J}_B$ of spin $B$ will inherit a negative sign.
To avoid misunderstanding, we denote $-\hat{J}_B=J_B$, $\hat{J}_{A}=J_{A}$ and $\hat{J}_{C}=J_{C}$ hereafter.
The model Hamiltonian (\ref{eq:h-start}) is then recast as
\begin{equation}
	H = \mathcal{H}_0 + \mathcal{H}_{\rm SC} \,,\label{eq:h-single-charge}
\end{equation}
where
\begin{align}
	\mathcal{H}_0&= \sum_{j,\sigma} \big[\Gamma_{\sigma} c_{j\sigma}^\dag c_{j\sigma}
	- (J_{\sigma}c_{j+1,\sigma}^\dag c_{j,\sigma} + \rm H.c.)\big]\,, \label{eq:h-single-charge-rotated-h-0}\\
	\mathcal{H}_{\rm SC}&= \sum_{j} i( \alpha_{AB} c_{j+1,A}^\dag c_{j,B} - \alpha_{AB} c_{j-1,A}^\dag c_{j,B} \notag\\
	& +\alpha_{CB} c_{j+1,C}^\dag c_{j,B} - \alpha_{CB} c_{j-1,C}^\dag c_{j,B}) + \rm H.c.
	\label{eq:h-single-charge-rotated-h-sc}
\end{align}
The lattice model described by (\ref{eq:h-single-charge}) is illustrated by Fig. \ref{fig:single}.
Diagonalizing Hamiltonian (\ref{eq:h-single-charge}) can give the quasiparticle spectrum $E_\nu$ of the system, where $\nu$ denotes the index of the quasiparticle modes.

To analyze the physics governed by Hamiltonian (\ref{eq:h-single-charge}),
we transform it into the momentum-$k$ space.
By choosing the basis $\Psi_k=(c_{k,A},c_{k,B},c_{k,C})^T$, the Hamiltonian is written as
\begin{equation}
	H(k) = \begin{pmatrix}
		\xi_A(k)+\Gamma_A & \zeta_{AB}(k) & 0\\
		\zeta_{AB}(k) & \xi_B(k)+\Gamma_B & \zeta_{CB}(k) \\
		0 & \zeta_{CB}(k) & \xi_C(k)+\Gamma_C
	\end{pmatrix} \,, \label{eq:h-k-space}
\end{equation}
where $\xi_\sigma(k) = -2J_\sigma\cos(k/k_L)$ and $\zeta_{\tau=AB,CB}(k)=2\alpha_\tau \sin(k/k_L)$, and $k_L=\pi/\lambda_L$.
We can see that the elements of Eq. (\ref{eq:h-k-space}) are all real and the Hamiltonian preserves the $PT$-symmetry.
The order-parameter space of Hamiltonian (\ref{eq:h-k-space}) is described by $M_3 = O(3)/O(1)^3$,
and thereby the fundamental homotopy group of the system is expressed by the non-Abelian quaternion group:
$\pi_1(M_3)=Q_8$.

Since Hamiltonian (\ref{eq:h-k-space}) consists of three bands, its form in the parameter space can be recognized as
$H(k) = \mathcal{R}_k D \mathcal{R}_k^T$,
where $D$ is a 3$\times$3 diagonal matrix whose elements describe the flatted band energies,
and $\mathcal{R}_k$ serves as the othogonal rotation from the eigenstates of $D$ to those of $H(k)$.
The non-Abelian quaternion charges can be extracted by the Zak phases of $H(k)$'s eigenstates $\ket*{\Psi_k^{(n)}}$ ($n=1,2,3$) when $k$ swaps over the whole 1D Brillouin zones (BZ) \cite{Atala2013Dec-zak-phase-exp,Unal2020Jul-prl-zak-ref,Bouhon2020Sep-prb-zak-ref},
and the Zak phase of the $n$-th band is obtained by~\cite{Zak1989,Zheng2017Jan-zak-ref}
\begin{equation}
	\phi_{\rm Zak}^{(n)}= i \int_{\rm BZ} \bra*{\Psi_k^{(n)}}\partial_k\ket*{\Psi_k^{(n)}} d k.
\end{equation}
On one hand, Hamiltonian (\ref{eq:h-k-space}) is specified into two trivial conjugacy classes
if the phases of the $H(k)$'s eigenstates acquire zero (modulus $2\pi$) \cite{Vanderbilt2018Oct-berry-phase-book}.
These two trivial conjugacy classes correspond to the non-Abelian topological charges $\{1, -1\}$.
In physics, they can be depicted by systems with no coupling between any spins.
On the other hand, Hamiltonian (\ref{eq:h-k-space}) is specified into three nontrivial conjugacy classes
when $\mathcal{R}_k=e^{\frac{L_\eta}{2}k}$ describes the rotation along the $L_\eta~(\eta=x,y,z)$ axis.
Here $(L_\eta)_{ij}=-\epsilon_{\eta ij}$ and $\epsilon_{\eta ij}$ denotes the antisymmetric tensor.
In these cases, the Zak phases are $\pi$ (modulus to $2\pi$) for two bands and 0 for the third one.
They correspond to the charges $\{\pm i, \pm j, \pm k\}$,
and describe systems with coupling between particular spins because $L_\eta$ belong to skew matrices.
As the results, it is accessible to generate nontrivial charges via manipulating the coupling within different spins.
Our focus in this work is on engineering the nontrivial quaternion charges, and thus we consider the following cases:
(i) For Hamiltonian of charge $\pm i$ (or $\pm k$),
the lower (or higher) two bands host nonzero Zak phases, while the third band is trivial.
Hence, it can be engineered by introducing coupling from spin $B$ to $C$ (or $A$).
(ii) For Hamiltonian of charge $\pm j$,
The lowest and highest bands host nonzero Zak phases, while the third band is trivial.
Hence, it can be engineered by introducing coupling from spin $B$ to both $C$ and $A$.
We show the specifications of their engineering in Fig. \ref{fig:single}.

As for such a 1D system, topological edge modes, whose density distribution is located on lattice boundaries, are expected to emerge by connecting two bands with nontrivial Zak phases. Therefore, the system of different quaternion charges exhibits distinguished behaviors on its boundaries, due to the Zak phases of its three bands. Since the three elements of the eigenstates are real, we can parametrize and illustrate them on a unit sphere. In Fig. \ref{fig:spectra of single charge}, we plot the evolution of the eigenstates in the pseudospin space when $k$ swaps over the BZ. For a Hamiltonian of charge $\pm i$ (or $\pm k$), the eigenstates for the lowest (or highest) two bands acquire a $\pi$ phase when evolving in BZ, while it remains unchanged for the third band. Consequently, the topological edge modes reside in the gap of the lowest (or highest) two bands, as shown in Figs. \ref{fig:spectra of single charge}(c1) and \ref{fig:spectra of single charge}(c3). For a Hamiltonian of charge $\pm j$, the eigenstates for both the lowest and highest bands acquire the $\pi$ phase, while it returns to the initial state for the third band. Hence for a Hamiltonian of charge $\pm j$, the topological edge modes exist between the lowest and highest bands. They are either separately present in different gaps or totally merge into the third band, as shown in Fig. \ref{fig:spectra of single charge}(c2). This motivates us to investigate the interface physics of a junction connecting different quaternion charges, which can extract the interesting non-Abelian features.

\section{Non-Abelian Features}\label{sec:interface-modes}

\begin{figure}[t]
	\centering
	\includegraphics[width=0.48\textwidth]{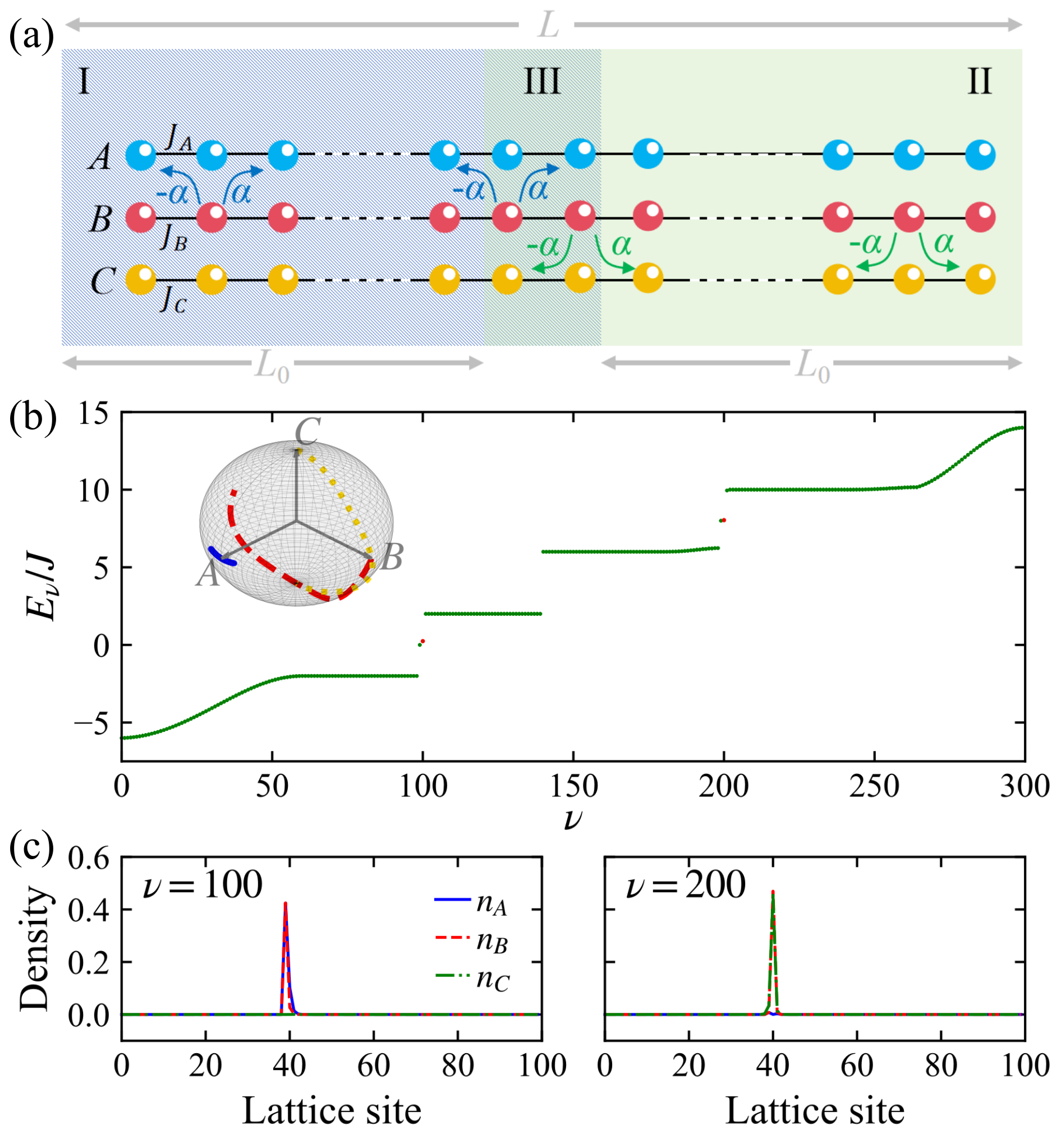}
	\caption{
		(a) Setups of the 1D Lattice model with two quaternion charges.
		Two external optical fields are applied to generate the coupling between different spins, but are spatially offset.
		The model consists of three sectors labeled as I, II, and III.
		In I (or II), only one coupling is present, engineering the quaternion charge $k$ (or $i$).
		In III as the overlapping area of the external optical fields, both two couplings are present.
		(b) The spectrum of the system composed of three sectors modeled in panel (a).
		The topological interface modes are marked by the red dots.
		The inner panel shows the evolution of eigenstates extracted from the subsystem described by III.
		(c) The density distribution of the interface modes on the domain wall between the sectors I and III.
		We set $L=100$, $L_0=40$,
		$(\Gamma_A^{(\rm I)},\Gamma_B^{(\rm I)},\Gamma_C^{(\rm I)})=(0, 0, 12J)$,
		and $(\Gamma_A^{(\rm II)},\Gamma_B^{(\rm II)},\Gamma_C^{(\rm II)})=(\Gamma_A^{(\rm III)},\Gamma_B^{(\rm III)},\Gamma_C^{(\rm III)})=(-4J, 8J, 8J)$.
		Other parameters in the sectors I and II are the same as Figs. \ref{fig:single}(c) and \ref{fig:single}(a), respectively.
	}
	\label{fig:spectrum of Combining-charges}
\end{figure}

Since the Hamiltonians of charges $\pm i$ and $\pm k$ are engineered by only one coupling in Fig. \ref{fig:single}, we focus on preparing the junction structure that connects the two quaternion charges. The setups are illustrated in Fig. \ref{fig:spectrum of Combining-charges}(a). We simultaneously introduce two external fields with spatial modulations as shown in Eq. (\ref{eq:coupling-pattern}) to generate Eq. (\ref{eq:h-single-charge-rotated-h-sc}), but with a spatial offset. Then, the 1D system can be regarded as a combination of three sectors as shown in Fig. \ref{fig:spectrum of Combining-charges}(a).

In sector I (II), only one external field is applied, and thereby atoms of spin $B$ are solely coupled with those of spin $A$ (or $C$), i.e., exhibiting a box-shaped SOC. In contrast, in the connecting area between I and II, atoms of spin $B$ are coupled to both spins $A$ and $C$, and we denote this area as sector III. For simplicity without loss of generality, we assume the length of the three sectors \{I, II, III\} along the chain is respectively set as \{$L_0$, $L_0$, $L-2L_0$\}, where $L$ stands for the lattice length. From Section \ref{sec:model}, we know that the Hamiltonian for such a 1D system is written as:
\begin{equation}
	H_{\rm junc} = \sum_{\eta={\rm I,II,III}}\mathcal{H}_{0}^{(\eta)} + \mathcal{H}_{\rm SC}^{(\eta)} \,, \label{eq:h-non-abelian-start}
\end{equation}
where the form of $\mathcal{H}_{0}^{(\rm \eta)}$ has been given in Eq. (\ref{eq:h-single-charge-rotated-h-0})
but with various parameter setups as $\Gamma_\sigma^{(\rm \eta)}$. $\mathcal{H}_{\rm SC}$ in different sectors are given as
\begin{align*}
	\mathcal{H}_{\rm SC}^{(\rm I)} &=\sum_{j} i\alpha(c_{j+1,A}^\dag c_{j,B} - c_{j-1,A}^\dag c_{j,B} )+\rm H.c. \\
	\mathcal{H}_{\rm SC}^{(\rm II)} &=\sum_{j} i\alpha(c_{j+1,C}^\dag c_{j,B} - c_{j-1,C}^\dag c_{j,B} )+\rm H.c.
\end{align*}
and $\mathcal{H}_{\rm SC}^{(\rm III)} =\mathcal{H}_{\rm SC}^{(\rm I)}+\mathcal{H}_{\rm SC}^{(\rm II)}$,
where we have denoted $\alpha_{AB}=\alpha_{CB}=\alpha$.
The subsystems described by Hamiltonian of sectors I and II host the quaternion charge $k$ and $i$, respectively.
As a result, topological interface modes are expected to exist on the domain walls of the sector III.
In Fig. \ref{fig:spectrum of Combining-charges}(b), we plot the spectrum of Hamiltonian (\ref{eq:h-non-abelian-start}).
Although in the overlapping area III, all the three spins are coupled,
we find the subsystem described by Hamiltonian of III indeed exhibits the behaviors of the charge $i$.
This can be seen in the inner panel of Fig. \ref{fig:spectrum of Combining-charges}(b), by extracting the evolution of eigenstates over BZ,
Therefore, as shown in Fig. \ref{fig:spectrum of Combining-charges}(c),
the soughted interface modes are located on the domain wall between I and III due to the different quaternion charges,
while are absent on the domain wall between III and II due to the same one.
We remark that since the spatial offset of SOC is tunable, it reveals the method for artificially manipulating and adjusting location of the interface modes.

Moreover, we find the interface modes are separately present in different gaps, yielding the interface modes are ascribed to a $j$ charge.
This result is the direct demonstration to the non-Abelian features of the quaternion charges,
because the behaviors of the interface modes correspond to the charge quotient $\Delta q=q_{\rm III}/q_{\rm I}$ of the two bulk sectors.
The non-Abelian bulk-edge correspondence is dominated by the noncommunicative multiplication relation of $i\cdot j = k$ for this case.
The demonstration to the other multiplication relations of quaternion charges is given in Appendixes \ref{sec:junction-j-and-k} and \ref{sec:junction-k-k}.

The aforementioned results are derived within an idealized framework of the box-shaped SOC associated with discontinuous boundaries.
In practical experiments, the SOC typically manifests as a Gaussian profile at the boundaries of the box.
We note that  this variation in the SOC profile does not alter the system's topological properties, provided that the band gaps remains open when SOC changes in the vicinity of the boundaries.
The detailed results are presented in Appendix \ref{sec:gaussian-beamer}.

\section{Experimental Implements}\label{sec:experiment}
Our proposal can be realized using current technology in ultracold atoms.
Here we use $^{40}$K atoms \cite{Li2020Jul-pra-K40-exp} as a concrete example.
The pseudospin states $A$, $B$, and $C$ are represented by the hyperfine levels of $\ket{F,m_F}=\ket{9/2,1/2}$, $\ket{9/2,-1/2}$, and $\ket{9/2,-3/2}$, respectively.
The atoms are loaded in a 1D optical lattice formed by counterpropagating lasers with wavelength $\lambda_{\rm Laser}=$ 1064 nm along the $x$ direction,
while are tightly bounded in the $y$ and $z$ directions.
Thus the lattice constant is given by $\lambda_L=\lambda_{\rm Laser}/2$.
The recoil energy of such an optical lattice is the recoil energy $E_R= h^2/(8m \lambda_L^2) \approx 2\pi\hbar\times$4.41kHz, which we choose as the energy unit below.
We set the lattice trap depth as $V_L=5.0E_R\approx$ 22.1kHz.
The corresponding hopping magnitude is $J=0.0658E_R$ \cite{Walters2013Apr-wannier-calculation}.
Since the on-site potential $\Gamma$ is spin dependent, it can be generated by the ac-Stark shift introduced via imposing the additional field.
To obtain the SOC strength $\alpha\approx 0.0644 E_R \approx 0.98J$ used in Figs. \ref{fig:spectra of single charge} and \ref{fig:spectrum of Combining-charges},
we tune the amplitude of the coupling fields in Eq. (\ref{eq:coupling-pattern}) as $\hat{\Omega}_{\sigma\sigma'}=5.5E_R\approx 24.3$kHz.
Since the topological modes in the band gaps exist as long as SOC is present,
$\hat{\Omega}_{\sigma\sigma'}$ can be further adjusted over a broad range without breaking the validity of the tight-binding approximation.

\begin{figure}[t]
	\centering
	\includegraphics[width=0.48\textwidth]{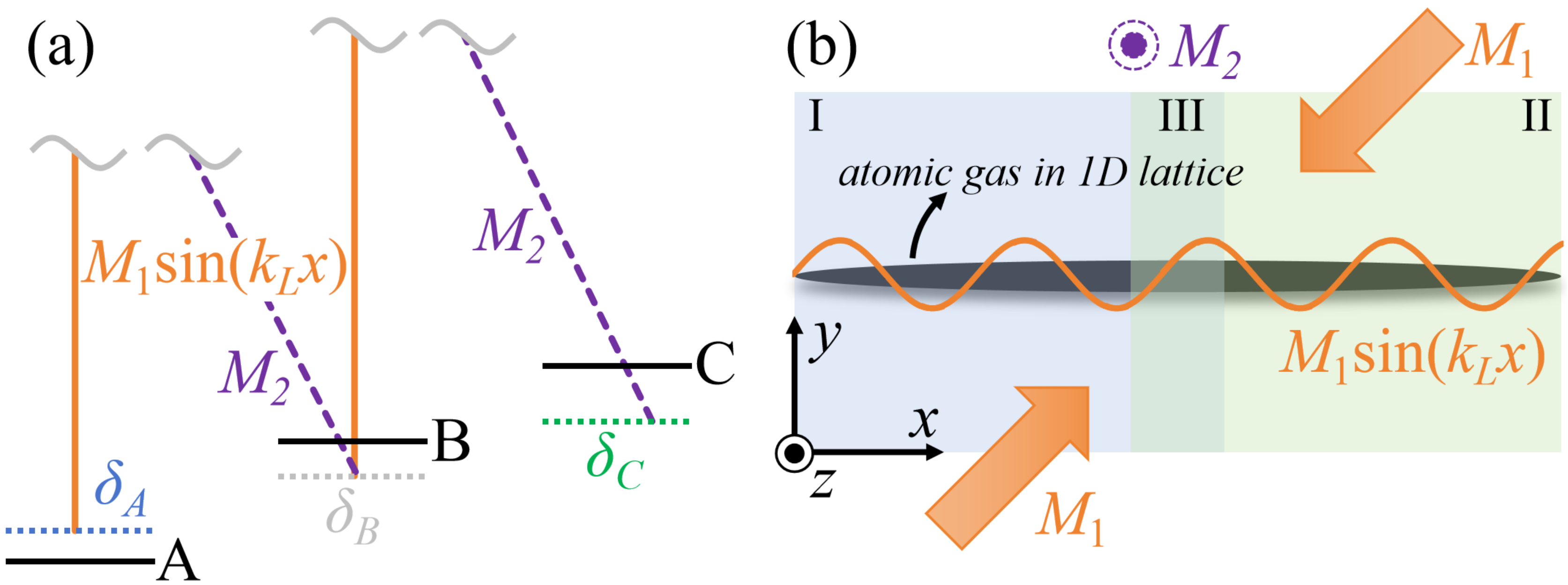}
	\caption{Experimental setups.
		(a) Sketch of the atomic transition for the SOC. The pseudospins are simultaneously coupled via two fields, denoted by $M_1$ (orange solid lines) and $M_2$ (purple dashed lines). The detuning of the spin-$\sigma$ state is represented by $\delta_\sigma$.
		(b) Setups for the junction structure that connects different quaternion charges.
		The atoms are loaded in the 1D optical lattice oriented along the $x$-axis.
		The field $M_1$ are produced by two counterpropagating lasers positioned in the $x$-$y$ plane, resulting to a sinusoidal mode of $\sin(k_Lx)$ (orange solid curve) when projecting onto the lattice.
		The field $M_2$ is aligned along the $z$ direction, perpendicular to the lattice.
		The detuning $\delta_{A}(\delta_{C})$ in panel (a) can be individually adjusted in the sector I (II) in panel (b).}
	\label{fig:exp-setup}
\end{figure}

Generally, SOC are produced by  two-photon Raman process that couples two pseudospins via an intermediate excited state \cite{Lin2011Mar-soc-exp}, as sketched in Fig. \ref{fig:exp-setup}(a).
In practice, we load the atoms into the optical lattice oriented along the $x$-axis, while the two optical fields that generate SOC are placed off the $x$-axis.
After adiabatically eliminating the excited states, it gives rise to a spatial modulation of Eq. (\ref{eq:coupling-pattern}) when projecting onto the 1D lattice.
Based on this scenario, the spatial dependence of SOC can be introduced by manipulating the detunings $\delta_{A}$, $\delta_{B}$, $\delta_{C}$ among three atomic states.
In particular, when one state is far detuned, the SOC is only present between the other two states.
For example as shown in Fig. \ref{fig:exp-setup}(b), we prepare that $\delta_A$ is far detuned in the sector I (resulting in charge $k$),
while $\delta_C$ is far detuned in the sector II (resulting in charge $i$).
This approach makes it attainable to create a junction structure that connects regions with different charges

Since the quaternion charges are determined by the Zak phases of the band structure,
the 1D atomic systems characterized by distinct charges can be identified through measurements of the Bloch oscillations, as reported in Ref. \cite{Atala2013Dec-zak-phase-exp}.
On the other hand, the evolution of the eigenstates for each band, from which the Zak phases are extracted as depicted in Figs. \ref{fig:spectra of single charge}(a1)-\ref{fig:spectra of single charge}(a3), can also be elucidated by analyzing the spin texture \cite{Wu2016Oct-soc-exp}.
The interface modes, as the eigenstate of the junction Hamiltonian, are then demonstrated in a similar way of the quantum state preparation in Ref. \cite{Braun2024NatPhys}.

\section{Conclusion}\label{sec:conclusion}

In summary, we have presented a scheme for synthesizing the non-Abelian charges in ultracold atoms.
The model  Hamiltonian for quaternion charges is established by implementing different spin-orbit couplings between two of the three pseudospin states. By adjusting the spatial misalignment of the spin-orbit coupling fields, a junction structure connecting various charges can be formed. This setup leads to the emergence of interface modes, exhibiting characteristics of non-Abelian bulk-edge correspondence. We believe that this approach offers a practical and versatile strategy for observing and manipulating the intriguing non-Abelian properties of the system.

\begin{acknowledgments}
	This work was supported by National Key Research and Development Program of China (Grant No. 2022YFA1405300), the National Natural Science Foundation of China (Grant No. 12074180), and the Science and Technology Program of Guangzhou (Grant No. 2024A04J0004).
\end{acknowledgments}

\appendix

\section{Case of the junction between $j$ and $k$}\label{sec:junction-j-and-k}

\begin{figure}[t]
	\centering
	\includegraphics[width=0.48\textwidth]{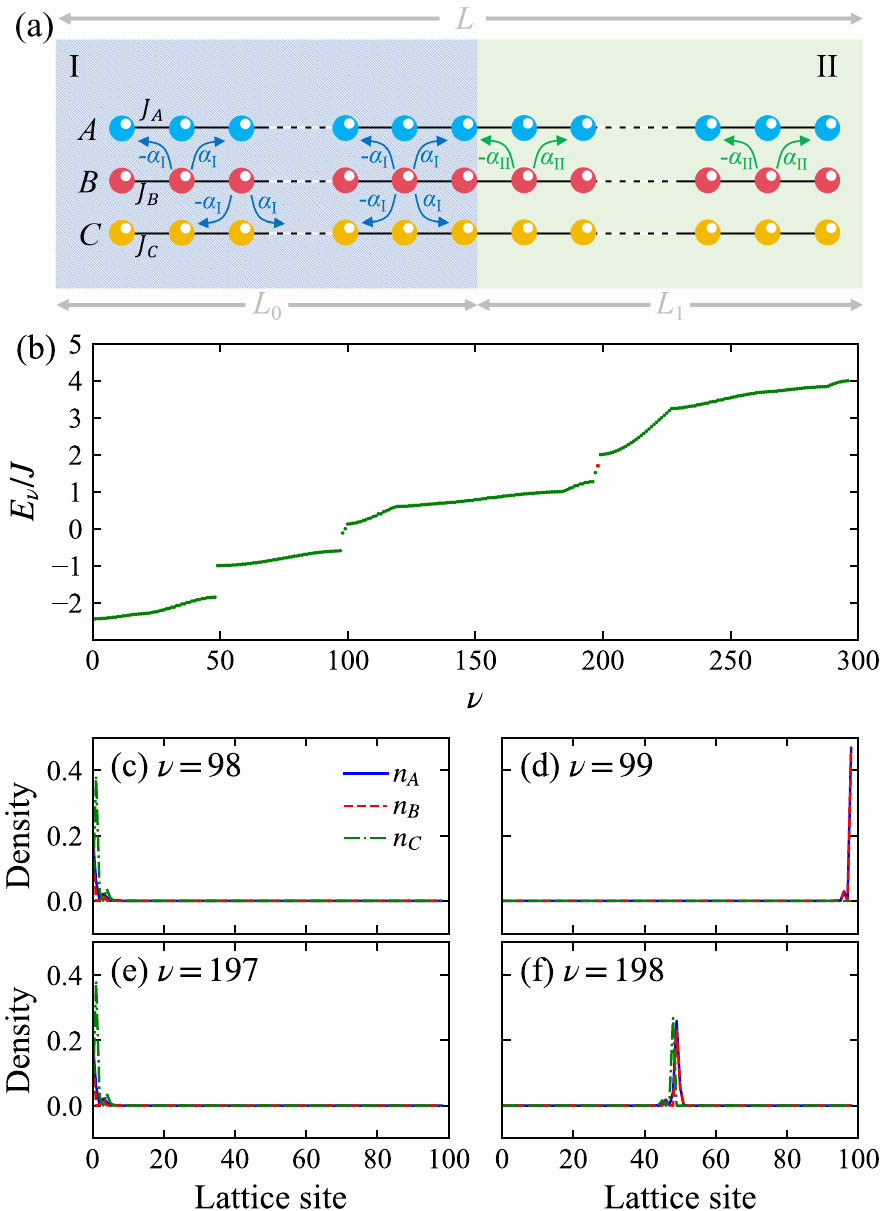}
	\caption{(a) Setups of the 1D Lattice model with two quaternion charges.
		The model consists of two sectors labeled as I and II.
		In II, only one coupling is present, engineering the quaternion charge $k$.
		In I, both two couplings are present, engineering the charge $j$.
		(b) The spectrum of the system composed of two sectors modeled in panel (a).
		The topological interface modes are marked by the red dots.
		(c)-(f) The density distribution of the edge states.
		We set $L=99$, $L_0=50$, $L_1=49$.
		Other parameters are as follows: $(\Gamma_A^{(\rm I)},\Gamma_B^{(\rm I)},\Gamma_C^{(\rm I)})=(2.8J, 2.8J, 2.8J)$;
		$(\Gamma_A^{(\rm II)},\Gamma_B^{(\rm II)},\Gamma_C^{(\rm II)})=(0, 0, 12J)$;
		$(J_A^{(\rm I)},J_B^{(\rm I)},J_C^{(\rm I)})=(J, J, 0)$;
		$(J_A^{(\rm II)},J_B^{(\rm II)},J_C^{(\rm II)})=(2J, 2J, 2J)$;
		$\alpha^{(\rm I)}=4J$; and $\alpha^{(\rm II)}=-1.2J$.
	}
	\label{fig:spectrum_j,k}
\end{figure}

In Sec. \ref{sec:interface-modes}, we discuss the junction system connecting the charges $k$ and $i$. To further illustrate the noncommunicative multiplication relation of the other charges,
hereafter we follow the setups of SOC presented in Fig. \ref{fig:spectrum of Combining-charges}(a) and demonstrate the case between $j$ and $k$
Since two external fields are applied for engineering the charge $j$,
the sector III can practically reduce to the interface site between the sectors I and II in this system,
as shown in Fig. \ref{fig:spectrum_j,k}(a).
The forms of the junction Hamiltonian $H_{\rm junc}$ and $\mathcal{H}_{0}^{(\rm \eta)}$ have been given by Eqs. (\ref{eq:h-non-abelian-start}) and (\ref{eq:h-single-charge-rotated-h-0}),
while $\mathcal{H}_{\rm SC}^{(\eta)}$ are instead given as
\begin{align*}
	\mathcal{H}_{\rm SC}^{(\rm I)} &=\sum_{j} i\alpha^{(\rm I)}(c_{j+1,A}^\dag c_{j,B} - c_{j-1,A}^\dag c_{j,B} \notag\\
	& - c_{j+1,C}^\dag c_{j,B} + c_{j-1,C}^\dag c_{j,B} )+\rm H.c. \\
	\mathcal{H}_{\rm SC}^{(\rm II)} &=\sum_{j} i\alpha^{(\rm II)}(c_{j+1,A}^\dag c_{j,B} - c_{j-1,A}^\dag c_{j,B} )+\rm H.c.
\end{align*}
where we have set $\alpha_{AB}^{(\rm I)}=-\alpha_{CB}^{(\rm I)}=\alpha^{(\rm I)}$, $\alpha_{AB}^{(\rm II)}=\alpha^{(\rm II)}$, $\alpha_{CB}^{(\rm II)}=0$.
The Hamiltonians of sectors I and II correspond to subsystems that host the charges $j$ and $k$, respectively. As the results, topological interface modes exist on the domain wall between I and II (i.e., at the interface site), as illustrated in Fig. \ref{fig:spectrum_j,k}(b). In the spectrum of this model, we find the interface mode exists between the highest two bands. Hence it can be ascribed to the charge $i$, and reveals the multiplication relation of $k\cdot i = j$.

\section{Case of the junction between $+k$ and $-k$}\label{sec:junction-k-k}

\begin{figure}[t]
	\centering
	\includegraphics[width=0.48\textwidth]{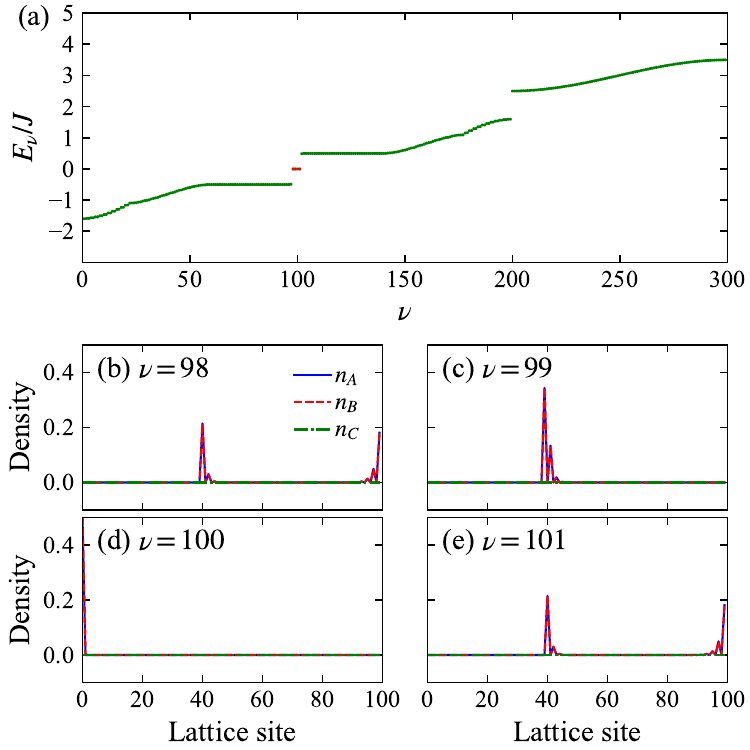}
	\caption{(a) The spectrum of the system composed of three sectors.
		The lattice setups are the same to Fig. \ref{fig:spectrum of Combining-charges}(a) but the junction structure connects charges $+k$ and $-k$ instead. The topological interface modes are marked by the red dots.
		(b-e) The density distribution of the edge states.
		We set $L=100$, $L_0=40$,
		and set $(\Gamma_A,\Gamma_B,\Gamma_C)=(0, 0, 12J)$ and
		$(J_A,J_B,J_C)=(J, J, J)$ for all three sectors.
		The SOC strengths are $\alpha^{(\rm I)}=-J$ and $\alpha^{(\rm II)}=3.2J$.
	}
	\label{fig:spectrum_k,-k}
\end{figure}

Here we investigate the junction between $+k$ and $-k$ as the example to show the results for charges of the same conjugate class.
In practice, such a junction structure can be constructed if one imposes a relative $\pi$ phase to the field [see Eq. (\ref{eq:coupling-pattern})] that generates SOC.
The forms of the junction Hamiltonian $H_{\rm junc}$ and $\mathcal{H}_{0}^{(\rm \eta)}$ have been given by Eqs. (\ref{eq:h-non-abelian-start}) and (\ref{eq:h-single-charge-rotated-h-0}),
while $\mathcal{H}_{\rm SC}^{(\eta)}$ are instead given as
\begin{equation*}
	\mathcal{H}_{\rm SC}^{(\eta ={\rm I,II})} =\sum_{j} i{\alpha}^{(\eta)}(c_{j+1,A}^\dag c_{j,B} - c_{j-1,A}^\dag c_{j,B} )+\rm H.c.
\end{equation*}
and $\mathcal{H}_{\rm SC}^{(\rm III)} =\mathcal{H}_{\rm SC}^{(\rm I)}+\mathcal{H}_{\rm SC}^{(\rm II)}$.
We have denoted ${\alpha}_{AB}^{(\rm I)}={\alpha}^{(\rm I)}$, ${\alpha}_{AB}^{(\rm II)}={\alpha}^{(\rm II)}$, ${\alpha}_{CB}^{(\rm I)}={\alpha}_{CB}^{(\rm II)}=0$.
Hamiltonian of the sector III belongs to the same charge with II if ${\alpha}^{(\rm I)}+{\alpha}^{(\rm II)}$ has the same sign with ${\alpha}^{(\rm II)}$.
Topological interface modes likewise exist on the domain wall between I and III as shown in Fig. \ref{fig:spectrum_k,-k}.
It is not difficult to observe that there are three edge states located at the interface, which can be ascribed to the $-1$ charge (see Appendix \ref{sec:charge-minus1}).
Therefore, this case further demonstrates the multiplication relation of  $k\cdot (-1) = -k$.

\section{Case of the charge $-1$}\label{sec:charge-minus1}

\begin{figure}[t]
	\centering
	\includegraphics[width=0.48\textwidth]{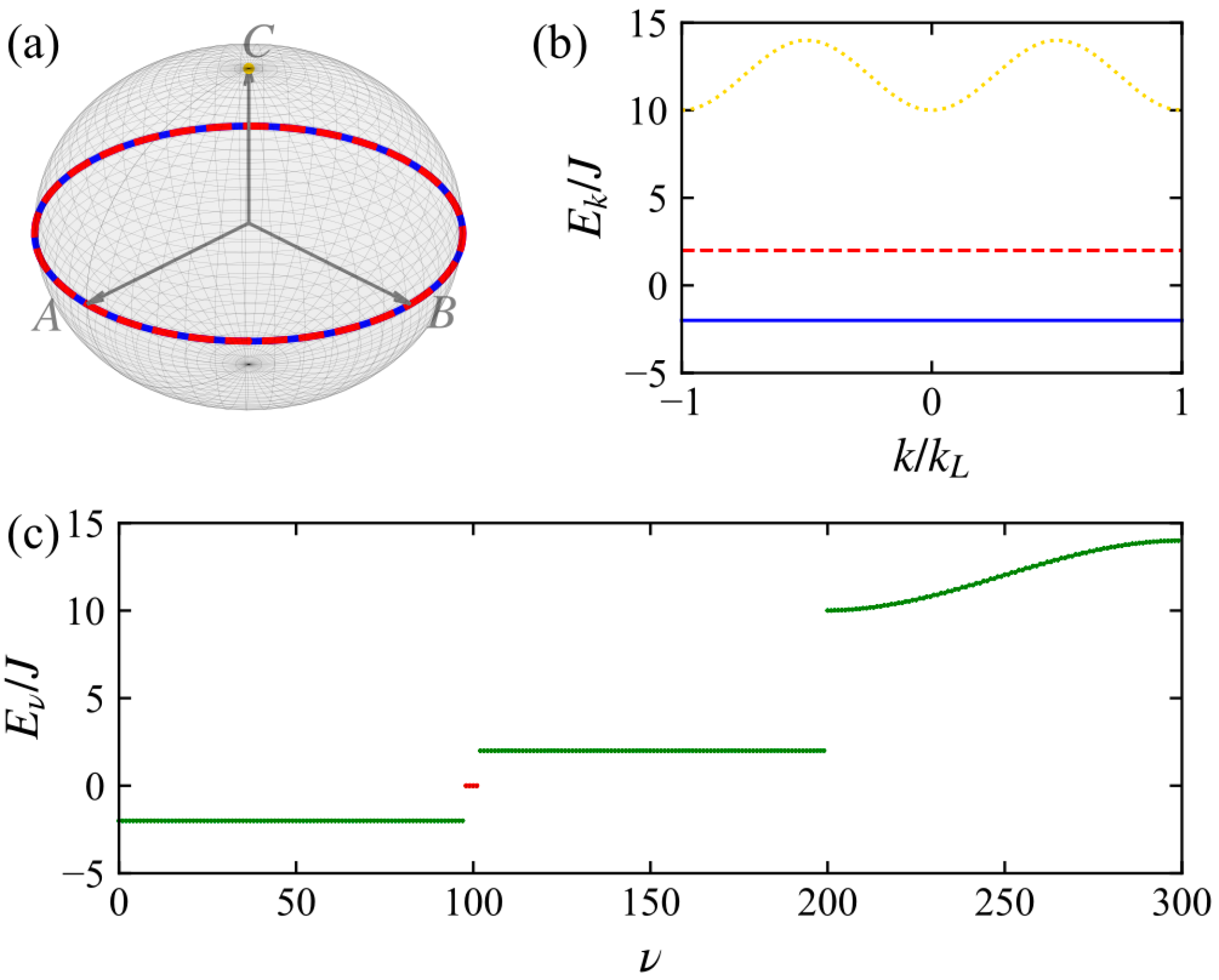}
	\caption{Band physics of the charge $-1$. Topological edge modes are marked by red dots.
		We set $J'_{A}=-J_{B}'=J'_{C}=J$, $(\Gamma_A,\Gamma_B,\Gamma_C)=(0, 0, 12J)$, and $(\alpha'_{AB},\alpha'_{CB})=(-J,0)$.}
	\label{fig:spectrum_-1}
\end{figure}

After discussing the nontrivial charges $\{i,j,k\}$, we finally study the Hamiltonian of the $-1$ charge.
We note that since the charge $-1$ corresponds to the system in which two of the three bands host a Zak phase of $2\pi$, engineering such a system requires the next-NN hopping as well as SOC, which are still challenging for current techniques in ultracold atoms.

The Hamiltonian that belongs to the charge $-1$ can be given as
\begin{equation}
	H_{-1} = \mathcal{H}_{0}' + \mathcal{H}_{\rm SC}' \,,\label{eq:h-single-charge_-1}
\end{equation}
where
\begin{equation}
	\mathcal{H}_{0}'= \sum_{j,\sigma} \Gamma_{\sigma} c_{j\sigma}^\dag c_{j\sigma}
	-(J'_{\sigma}c_{j+2,\sigma}^\dag c_{j,\sigma} + \rm H.c.) \label{eq:h-single-charge-rotated-h-0_-1}
\end{equation}
and
\begin{align}
	\mathcal{H}_{\rm SC}'&= \sum_{j} i( \alpha'_{AB} c_{j+2,A}^\dag c_{j,B} - \alpha'_{AB} c_{j-2,A}^\dag c_{j,B} \,,\notag\\
	& +\alpha'_{CB} c_{j+2,C}^\dag c_{j,B} - \alpha'_{CB} c_{j-2,C}^\dag c_{j,B}) + \rm H.c.
	\label{eq:h-single-charge-rotated-h-sc_-1}
\end{align}
We transform it into the momentum space, and the Hamiltonian is written as
\begin{equation}
	H_{-1}(k) = \begin{pmatrix}
		\xi'_{A}(k)+\Gamma_A & \zeta'_{AB}(k) & 0\\
		\zeta'_{AB}(k) & \xi'_{B}(k)+\Gamma_B & \zeta'_{CB}(k) \\
		0 & \zeta'_{CB}(k) & \xi'_{C}(k)+\Gamma_C
	\end{pmatrix} \,, \label{eq:h-k-space_-1}
\end{equation}
where $\xi'_{\sigma}(k) = -2J'_{\sigma}\cos(2k/k_L)$ and $\zeta'_{\tau=AB,CB}(k)=2\alpha'_{\tau} \sin(2k/k_L)$.
The band physics of the charge $-1$ are shown in Fig. \ref{fig:spectrum_-1}. For Hamiltonian of charge $-1$, the eigenstates for the lowest two bands acquire a 2$\pi$ phase when evolving in BZ, with the third band unchanged. Hence the edge states are located in the gap of the lowest two bands, as shown in Fig. \ref{fig:spectrum_-1}(c). Unlike the charge $\pm k$, Hamiltonian (\ref{eq:h-single-charge_-1}) of the charge $-1$ hosts the fickle edge states with beyond-two-fold degeneracy, which depend on the details of the system setups \cite{Guo2021Jun-non-abelian-charges-in-photonics}.

\section{Case of the box-shaped SOC with Gaussian boundaries} \label{sec:gaussian-beamer}

\begin{figure}[t]
	\centering
	\includegraphics[width=0.48\textwidth]{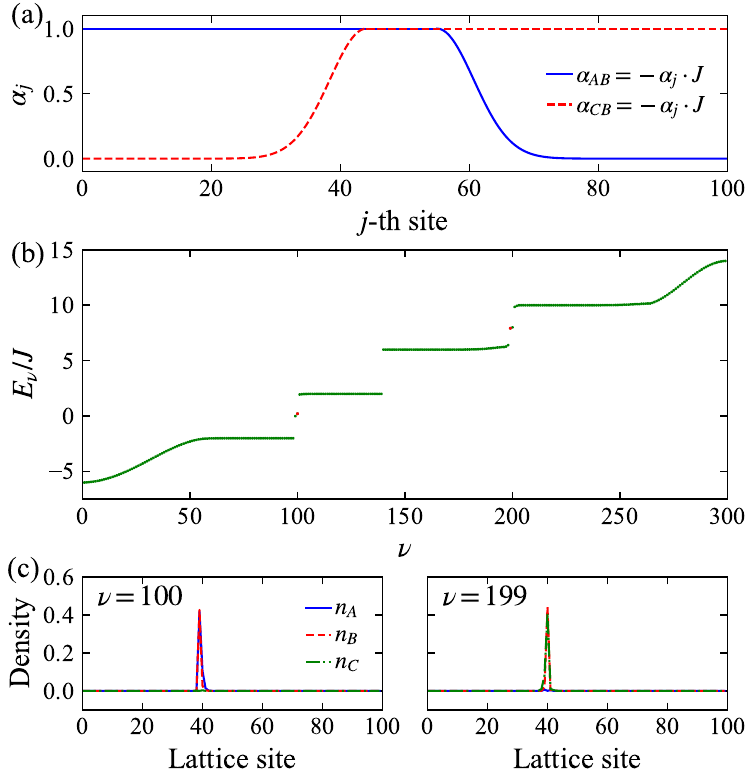}
	\caption{(a) Spatial dependence of SOC for the lattice model.
		(b) The spectrum of the system in Fig. \ref{fig:spectrum of Combining-charges} with the spatially dependent SOC of (a).
		The topological interface modes are marked by the red dots.
		(c) The density distribution of the interface states. The parameters are set as the same with Fig. \ref{fig:spectrum of Combining-charges}.}
	\label{fig:gaussian-beamer}
\end{figure}

In practical experiments involving the box-shaped optical fields \cite{Navon2021NatPhys,Busley2022Science,Leonard2023Nature},
the boundaries of the box usually render to a Gaussian-like profile.
Here we reperform the calculation of Fig. \ref{fig:spectrum of Combining-charges} by assuming the SOC changes spatially as shown in Fig. \ref{fig:gaussian-beamer}(a).
We find the topological interface modes are robust under the induced Gaussian profile, as seen in Figs. \ref{fig:gaussian-beamer}(b) and \ref{fig:gaussian-beamer}(c).
It indicates that the quaternion charges are topologically invariant under the adiabatic deformation of SOC without closing the band gaps.

%----------------------------------------------------------------------------------------
\vfill
\bibliographystyle{apsrev4-1}
\bibliography{ref}
\end{document}